\documentclass[conference]{IEEEtran}
\IEEEoverridecommandlockouts
\usepackage{cite}
\usepackage{amsmath,amssymb,amsfonts}
\usepackage{mathtools}
\usepackage{algorithm}
\usepackage{algpseudocode}

\usepackage{graphicx}
\usepackage{textcomp}
\usepackage{xcolor}
\def\BibTeX{{\rm B\kern-.05em{\sc i\kern-.025em b}\kern-.08em
    T\kern-.1667em\lower.7ex\hbox{E}\kern-.125emX}}

\makeatletter
\def\endthebibliography{%
	\def\@noitemerr{\@latex@warning{Empty `thebibliography' environment}}%
	\endlist
}
\makeatother

\usepackage{amsthm}
\newtheorem{definition}{Definition}

\makeatletter
\newcommand{\removelatexerror}{\let\@latex@error\@gobble}
\makeatother

\usepackage{todonotes}
\usepackage{graphicx}
\usepackage{hyperref}
\hypersetup{colorlinks}
\usepackage{siunitx}
\usepackage{tikz}
\usepackage{pgfplots}
\usepgfplotslibrary{patchplots,groupplots,fillbetween}
\pgfplotsset{compat=newest}
\usetikzlibrary{spy,arrows,positioning,shapes,shapes.geometric,external,shadows,babel,calc,math,plotmarks,intersections,arrows,decorations.markings}
 
\ifCLASSOPTIONcompsoc
\usepackage[caption=false, font=normalsize, labelfont=sf, textfont=sf]{subfig}
\else
\usepackage[caption=false, font=footnotesize]{subfig}
\fi

\newcommand{\orcidicon}[1]{\href{https://orcid.org/#1}{\includegraphics[height=\fontcharht\font`\B]{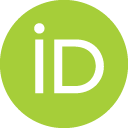}}}

\newcommand\copyrighttext{%
	\footnotesize \copyright\,2022 IEEE. Personal use of this material is permitted. Permission from IEEE must be obtained for all other uses, in any current or future media, including reprinting/republishing this material for advertising or promotional purposes, creating new collective works, for resale or redistribution to servers or lists, or reuse of any copyrighted component of this work in other works.}%
\newcommand\copyrightnotice{%
	\begin{tikzpicture}[remember picture,overlay]%
	\node[anchor=south,yshift=10pt] at (current page.south) {\fbox{\parbox{\dimexpr\textwidth-2cm}{\copyrighttext}}};%
	\end{tikzpicture}%
	\vspace{-10pt}%
}

\begin{document}

    \title{
    	Self-Assessment for Single-Object Tracking in Clutter Using Subjective Logic
        \thanks{
    	    This research is accomplished within the project SecForCARs (grant number 16KIS0795). 
    	    We acknowledge the financial support for the project by the Federal Ministry of Education and Research of Germany (BMBF).}
    }
    
    \author{
        \IEEEauthorblockN{
            Thomas Griebel\,\orcidicon{0000-0001-6521-3013},
            Johannes M\"uller\,\orcidicon{0000-0001-9286-0937}, 
            Paul Geisler,
            Charlotte Hermann\,\orcidicon{0000-0003-2770-5644},\\
            Martin Herrmann\,\orcidicon{0000-0002-7953-2354}, 
            Michael Buchholz\,\orcidicon{0000-0001-5973-0794}, and 
            Klaus Dietmayer\,\orcidicon{0000-0002-1651-014X}}
        \IEEEauthorblockA{
            \textit{Institute of Measurement, Control} \\
            \textit{and Microtechnology, Ulm University}\\
            89081 Ulm, Germany \\
            \{thomas.griebel, johannes-christian.mueller, paul.geisler, charlotte.hermann, \\
            martin.herrmann, michael.buchholz, klaus.dietmayer\}@uni-ulm.de
        }
    }
    
    \maketitle
    \copyrightnotice
    
    \begin{abstract}
        Reliable tracking algorithms are essential for automated driving.
        However, the existing consistency measures are not sufficient to meet the increasing safety demands in the automotive sector.
        Therefore, this work presents a novel method for self-assessment of single-object tracking in clutter based on Kalman filtering and subjective logic.
        A key feature of the approach is that it additionally provides a measure of the collected statistical evidence in its online reliability scores.
        In this way, various aspects of reliability, such as the correctness of the assumed measurement noise, detection probability, and clutter rate, can be monitored in addition to the overall assessment based on the available evidence.
        Here, we present a mathematical derivation of the reference distribution used in our self-assessment module for our studied problem.
        Moreover, we introduce a formula that describes how a threshold should be chosen for the degree of conflict, the subjective logic comparison measure used for the reliability decision making.
        Our approach is evaluated in a challenging simulation scenario designed to model adverse weather conditions.
        The simulations show that our method can significantly improve the reliability checking of single-object tracking in clutter in several aspects.
    \end{abstract}
    
    \begin{IEEEkeywords}
        self-assessment, tracking, subjective logic, monitoring, Kalman filtering, nearest neighbor association
    \end{IEEEkeywords}
    
    \section{Introduction}

    Reliable object tracking is a keystone to safe automated driving.
    For example, for adaptive cruise control systems, automated vehicles have to track the vehicle driving ahead to safely follow the respective road user.
    While simple consistency measures, like the \textit{normalized innovation squared} (NIS) and the \textit{normalized estimation error squared} (NEES)~\cite{bar1988tracking, bar2001estimation}, have been used for a long time in single-object tracking (SOT), these methods can only provide limited insights into the reliability of the tracking.
    However, particularly in the light of the standard ISO $21448$ \textit{safety of the intended functionality} (SOTIF)~\cite{iso201921448}, reliability beyond functional safety recently gained increased attention in the automotive sector.

    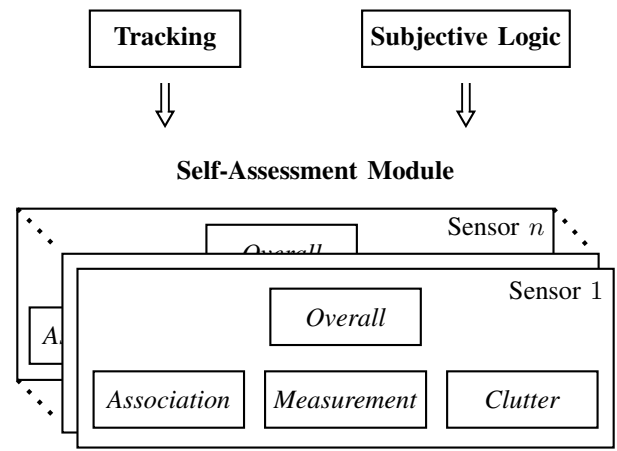
\begin{figure}[!t]
        \centering
     	\begin{tikzpicture}[>=latex']
     	    \tikzset{block/.style= {draw, rectangle, align=center,minimum width=2.0cm,minimum height=0.75cm},
            }
            \tikzstyle{vecArrow} = [thick, decoration={markings,mark=at position
            1 with {\arrow[thick]{open triangle 60}}},
            double distance=2.0pt, shorten >= 6.5pt,
            preaction = {decorate},
            postaction = {draw,line width=2.0pt, white,shorten >= 4.5pt}]
            \tikzstyle{innerWhite} = [thick, white,line width=2.0.0pt, shorten >= 4.5pt]
            
            \node (start) at (0,0) {\textbf{Self-Assessment Module}};
            \node [block, thick, above of=start, xshift=2cm, yshift=0.75cm] (sl){\textbf{Subjective Logic}};
            \node [block, thick, above of=start, xshift=-2cm, yshift=0.75cm] (track){\textbf{Tracking}};
            \node [block, thick, below of=start, xshift=0.2cm, yshift=-0.75cm] (sa-overall){\textit{Overall}};
            \node [block, thick, below of=sa-overall, xshift=-2.35cm, yshift=-0.1cm] (sa-asso){\textit{Association}};
            \node [block, thick, below of=sa-overall, xshift=0cm, yshift=-0.1cm] (sa-meas){\textit{Measurement}};
            \node [block, thick, below of=sa-overall, xshift=2.35cm, yshift=-0.1cm] (sa-clutter){\textit{Clutter}};
            \node [block, thick, below of=start, xshift=-0.45cm, yshift=-0.1cm] (sa02-overall){\textit{Overall}};
            \node [block, thick, below of=sa02-overall, xshift=-2.35cm, yshift=-0.1cm] (sa02-asso){\textit{Association}};
        		
     		\draw[vecArrow] ($(sl.south)+(0.0,-0.2)$) to ($(sl.south)+(0.0,-0.8)$);
     		\draw[vecArrow] ($(track.south)+(0.0,-0.2)$) to ($(track.south)+(0.0,-0.8)$);

     		\draw[thick] ($(sa-asso.north west)+(-0.8,+1.95)$) rectangle ($(sa-clutter.south east)+(-0.4,+0.45)$);
     		\draw[thick,fill=white,fill opacity=1.0] ($(sa-asso.north west)+(-0.2,+1.35)$) rectangle ($(sa-clutter.south east)+(0.2,-0.25)$);
     		\draw[thick,fill=white,fill opacity=1.0] ($(sa-asso.north west)+(0.0,+1.15)$) rectangle ($(sa-clutter.south east)+(0.4,-0.45)$);
     		\draw [ultra thick, loosely dotted] ($(sa-asso.north west)+(-0.8,+1.95)$) -- ($(sa-asso.north west)+(-0.25,+1.4)$);
     		\draw [ultra thick, loosely dotted] ($(sa-clutter.north east)+(-0.4,+1.95)$) -- ($(sa-clutter.north east)+(+0.1,+1.4)$);
     		\draw [ultra thick, loosely dotted] ($(sa-asso.south west)+(-0.8,+0.45)$) -- ($(sa-asso.south west)+(-0.25,-0.1)$);
     		
        		
            \node [block, thick, below of=start, xshift=0.4cm, yshift=-0.95cm] (sa-overall){\textit{Overall}};
            \node [block, thick, below of=sa-overall, xshift=-2.35cm, yshift=-0.1cm] (sa-asso){\textit{Association}};
            \node [block, thick, below of=sa-overall, xshift=0cm, yshift=-0.1cm] (sa-meas){\textit{Measurement}};
            \node [block, thick, below of=sa-overall, xshift=2.35cm, yshift=-0.1cm] (sa-clutter){\textit{Clutter}};
            \node [above of=sa-overall, xshift=2.8cm, yshift=-0.65cm] (sa-sensor01){Sensor $1$};
            \node [above of=sa-overall, xshift=2.0cm, yshift=+0.2cm] (sa-sensor0n){Sensor $n$};
            
        \end{tikzpicture}
        \caption{Our proposed self-assessment module for single-object tracking in clutter is visually displayed.
        The self-assessment module builds on a tracking algorithm and uses the subjective logic theory to obtain self-assessment measures.
        Our self-assessment module monitors tracking assumptions regarding the association, measurement, and clutter, in addition to the overall tracking performance.
        \label{fig:concept-sa-sot-clutter}}
    \end{figure}
    In this work, we propose a novel self-assessment (SA) scheme for SOT in clutter with nearest neighbor association using subjective logic (SL)~\cite{josang2016subjective}; see Fig.~\ref{fig:concept-sa-sot-clutter}.
    Our approach can provide both an assessment of the overall situation and a detailed look at specific aspects of the reliable operation of the SOT algorithm.
    Namely, the assumptions about the measurement noise, the clutter rate, and the detection probability can be assessed separately.
    The SA module we developed is the first approach in literature so far that can monitor the overall performance and certain statistical assumptions of the tracking algorithm.
    Since our online SA scores additionally feature measures for the statistical uncertainty, our method allows an evidence-based weighting of the individual aspects when combining them into the overall measure.
    From a theoretical perspective, we provide a mathematical derivation of the reference distribution used in our SA module.
    Furthermore, a novel threshold calculation for the SL's comparison measure, the degree of conflict (DC), is presented. 
    Our SA module uses the dynamically calculated threshold for the reliability decision-making.
    We evaluate our approach in a challenging simulation scenario motivated by adverse weather conditions, showing its capability to detect violations of the respective assumptions.
    
    Our main contributions can be summarized as follows:
    \begin{itemize}
        \item a novel overall concept and whole module for SA in SOT in clutter with an overall and component analysis using SL,
        \item online SA scores that additionally feature measures for the statistical uncertainty,
        \item a theoretical derivation of the reference distribution for our SA module, and
        \item a novel threshold calculation for the SL comparison measure DC.
    \end{itemize}

    \section{Related Work} \label{section:related-work}
    
    The NIS and the NEES are the classical consistency measures in Kalman filtering~\cite{bar1988tracking, bar2001estimation}.
    While the NEES requires ground truth data, the NIS can work in online applications without further knowledge of ground truth data.
    For the classical assumptions in Kalman filtering, i.e., all signals and probability densities are Gaussian distributed and the models are linear, both consistency measures follow a $\chi^2$ distribution.  
    Over the last decades, starting from the NIS, further consistency and SA measures for online applications have been investigated.
    
    In~\cite{gibbs2013new}, three tests for inconsistencies are proposed in the context of Kalman filtering.
    A fixed-interval smoother is used to improve a basic residual test to detect measurement outliers and filter model inconsistencies.
    Piché shows some relations of the NIS to other model criticism procedures in~\cite{piche2016online}.
    Consequently, a new test similar to the NIS is proposed, which matches a posterior predictive Bayesian $p$-test.
    In~\cite{gamse2014statistical}, Gamse et al.~present a comprehensive evaluation of a Kalman filter.
    This evaluation is based on statistical tests and several indicators of inner confidence like controllability and observability, the determinant of the state transition matrix, and properties of the Kalman gain.
	
    In~\cite{scalzo2009adaptive}, following the concept of the NIS, consistency checks for nonlinear filters in single-target tracking are considered.
    The paper presents an adaption procedure based on NIS-like tests for an extended and unscented Kalman filter and a particle filter.
    In~\cite{ivanov2014evaluating}, Ivanov et al.~propose the normalized deviation squared (NDS), which is claimed to be useful for nonlinear systems and sub-optimal filters. 
    However, the test is claimed to be more effective in offline consistency evaluation and mostly detects inconsistencies of not sufficiently large noise parameters.
    For the generalization of the NIS towards multi-target tracking, Mahler introduces a more general approach based on the NIS and calls it a divergence detector in~\cite{mahler2013divergence}.
    This includes the concept of the generalized NIS (GNIS), which is expanded to the multi-target generalized NIS (MGNIS).
    Explicit formulas of these divergence detectors are derived for the cardinalized probability hypothesis density filter and the probability hypothesis density filter.
    Based on~\cite{mahler2013divergence}, Reuter et al.~\cite{reuter2013divergence} derive the MGNIS for the $\delta$-generalized labeled multi-Bernoulli filter and develop an approximation, the so-called approximate multi-target NIS (AMNIS).
    Additionally, \cite{reuter2013divergence}~states that the MGNIS has its difficulties with the clutter component.
    In~\cite{stubler2017consistency}, consistency checks in a feature-based random-set Monte-Carlo localization are proposed closely related to the NIS.
    In particular, \cite{stubler2017consistency}~checks the consistency of several components in the measurement model as, e.g., spatial uncertainty, clutter rate, and missed detection, separate from each other. 
    
    Connecting SL with temporal filtering, our previous work~\cite{muller2019subjective} presents an identification algorithm of Markov chains using SL.
    By the use of the SL-based identification algorithm, an additional explicit reliability measure based on statistical uncertainty is obtained in contrast to classical methods.
    In~\cite{griebel2020kalman}, we investigate the connection between SL and Kalman filtering.
    For the first time in literature, we proposed a novel SA measure for Kalman filtering that additionally includes an explicit measure for the statistical uncertainty.
	
    This paper builds on our previous work and proposes the first extensive SA module for SOT in clutter with a nearest neighbor association in literature.
    We do obtain not only an overall assessment score by our SA module but also an assessment score for each reliability aspect of the tracking algorithm, i.e., the correctness of the assumed measurement noise, detection probability, and clutter rate.
    Here, we still consider the dependencies between the individual reliability aspects.

    \section{Fundamentals} \label{section:fundamentals}
    
    This section summarizes the conceptual formulation of SOT in clutter, which is mainly based on~\cite{challa2011fundamentals}.
    Furthermore, we describe the relevant mathematical concepts of SL from~\cite{josang2016subjective}.
    
    \subsection{Single-Object Tracking in Clutter}

    SOT in clutter is a special case of multi-object tracking where exactly one object is present.
    In addition to classical Kalman filtering, there are new challenging aspects. 
    Besides noisy measurements and estimating time varying states, we have to deal with missed detections, clutter detections, and unknown data associations.
    In fact, for SOT in clutter, the key changes compared to classical Kalman filtering are in the measurement model, while the concepts of prediction and update and object dynamics mostly stay the same.
    To this end, we focus on the measurement likelihood in the following.
    
    At each time step $k \in \mathbb{N}$, we have a set of measurements 
    $Z_k = \{z^1, \ldots, z^{m_k} \}$ with $z^i \in \mathbb{R}^{m_z}$, $i=1, \ldots, m_k$.
    This means that $m_z \in \mathbb{N}$ is the dimension of the measurement space for each individual measurement, and $m_k \in \mathbb{N}_0$ is the number of measurements at time step $k \in \mathbb{N}$.
    The estimated object state is modeled by $x_k \in \mathbb{R}^{n}$ with state dimension $n \in \mathbb{N}$.
    In Kalman filtering, $x_k$ is modeled by an $n$-dimensional multivariate Gaussian distribution with mean $\hat{x}_k \in \mathbb{R}^{n}$ and covariance matrix $P_k \in \mathbb{R}^{n \times n}$.
    
    Using Bayes's rule, the posterior density in time step $k \in \mathbb{N}$ is
    \begin{align} \label{eqn:posterior_density}
        p\left(x_k | Z_{1:k}\right) &= \frac{p\left(Z_k | x_k\right) p\left(x_k | Z_{1:k-1}\right)}{p\left(Z_k | Z_{1:k-1}\right)} ,
    \end{align}  
    where 
    $Z_{1:k} = \left(Z_1, \ldots, Z_k\right)$ is the time sequence of measurements,
    $p(Z_k | x_k)$ the measurement likelihood,
    $p(x_k | Z_{1:k-1})$ the prediction density,
    and $p(Z_k | Z_{1:k-1})$ the normalizing factor.
    The normalization term is given by
    \begin{align} \label{eqn:normalization_factor}
        p\left(Z_k | Z_{1:k-1}\right) &= \int p\left(Z_k | x_k\right) p\left(x_{k} | Z_{1:k-1}\right) d x_{k} \, .
    \end{align} 
    The prediction density is obtained using the Chapman-Kolmogorov equation
    \begin{align} \label{eqn:prediction_density}
        p\left(x_k | Z_{1:k-1}\right) &= \int p\left(x_{k-1} | Z_{1:k-1}\right) p\left(x_{k} | x_{k-1}\right) d x_{k-1} \, .
    \end{align}
    The measurement model is described in detail below.
    First, $p_D(x_k) \in [0,1]$ models the probability that a sensor detects an object at time step $k \in \mathbb{N}$, called the detection probability.
    Consequently, $1-p_D(x_k)$ is the probability that an object is undetected by the sensor.
    This means the object detection process of a sensor is Bernoulli distributed with probability $p_D(x_k)$.
    If detected, the measurement follows a single-object spatial likelihood function $g_k (z | x_k)$, where each measurement $z \in \mathbb{R}^{m_z}$ is generated by at most one object.
    Moreover, clutter measurements are modeled using a false alarm process where the number of clutter measurements $m_k^c \in \mathbb{N}_0$ is assumed to follow a Poisson distribution with the expected number $\bar{\lambda}_c \in (0, \infty)$, i.e., $m_k^c \sim \text{Poi} \left( \bar{\lambda}_c \right)$.
    All clutter measurements are independent and identically distributed with spatial distribution $\lambda_c (z)$. 
    Typically, it is assumed that $\lambda_c (z)$ is uniformly distributed over the sensor field of view (FOV). 
    To take the unknown data association of all measurements $Z_k$ at time step $k \in \mathbb{N}$ into account, we introduce $\theta_k \in \mathbb{N}_0$.
    With $\theta_k = 0$, we denote the hypothesis that all measurements are clutter, i.e., a missed detection.
    For $\theta_k \in  \{1, \ldots, m_k\}$, we denote the association hypotheses that measurement $z^{\theta_k}$ is an object measurement and the rest of the measurements, i.e., $z^i$ for all $i=1, \ldots, m_k$ with $i \neq \theta_k$, are clutter measurements.
    Putting all components together, we formulate the complete measurement likelihood using the law of total probability as
    \begin{align} 
        p(Z_k | x_k) =& \sum_{\theta_k=0}^{m_k} p(Z_k, m_k, \theta_k | x_k) \\
        =& \sum_{\theta_k=0}^{m_k} p(Z_k | m_k, \theta_k, x_k) p(\theta_k, m_k | x_k) \label{eqn:measurement-likelihood-general} \\
        =& \left[ (1-p_D(x_k)) +  p_D(x_k) \sum_{\theta_k=1}^{m_k} \frac{g_k (z^{\theta_k} | x_k)}{\lambda_c \left( z^{\theta_k} \right)} \right] \nonumber\\
        & \times \frac{\exp{(-\bar{\lambda}_c})}{m_k!} \prod_{i=1}^{m_k} \lambda_c (z^i) \, , \label{eqn:measurment-likelihood}
    \end{align}  
    where we sum over all possible data association hypotheses $\theta_k = 0, \ldots, m_k$.
    
    Substituting the prediction density~\eqref{eqn:prediction_density}, the normalization factor~\eqref{eqn:normalization_factor}, and the measurement likelihood~\eqref{eqn:measurement-likelihood-general} into the posterior density~\eqref{eqn:posterior_density}, we end up with the optimal Bayesian filter for SOT in clutter. 
    One possible approximation of this optimal recursion is the nearest neighbor association algorithm.
    The reason we need approximations of the optimal Bayesian filter is that the number of hypotheses grows with time as 
    $\prod_{j=1}^{k} \left( m_j +1 \right)$.
    As the number of time steps increases, it becomes more and more intractable to compute the exact solution 
    $p\left(x_k | Z_{1:k}\right)$ 
    of the optimal Bayesian filter. 
    The nearest neighbor algorithm assigns the measurement closest to the predicted measurement in terms of a statistical distance to the object.
    All other measurements are discarded and considered as clutter.
    For more information on the nearest neighbor algorithm and other association algorithms, see~\cite{challa2011fundamentals}.

    \subsection{Subjective Logic}
    
    In this section, several components of SL are mathematically presented, which rest on the work of J{\o}sang in~\cite{josang2016subjective}.
    The central element of SL is the opinion representation.
    \begin{definition}[Multinomial Opinion] \label{def:multinomial-opinion}
    	Let $X \in \mathbb{X}$ be a random variable of the finite domain $\mathbb{X}$ with cardinality $W = |\mathbb{X}| \geq 2$. 
    	A multinomial opinion is an ordered triple $\omega_X = (\boldsymbol{b}_X,u_X,\boldsymbol{a}_X)$ with
    	\begin{subequations}
    		\begin{align}
    		    \boldsymbol{a}_X(x) : \mathbb{X} \mapsto [0,1], \qquad 1 &= \sum\limits_{x \in \mathbb{X}} \boldsymbol{a}_X (x) \, ,\\
    		    \boldsymbol{b}_X(x) : \mathbb{X} \mapsto [0,1], \qquad 1 &= u_X + \sum\limits_{x \in \mathbb{X}} \boldsymbol{b}_X (x)\, .
    		\end{align}
    	\end{subequations}
    	In detail, opinions consist of the belief mass distribution $\boldsymbol{b}_X$ over $\mathbb{X}$,
    	the uncertainty mass $u_X \in [0,1]$ representing the lack of evidence,
    	and the base rate distribution $\boldsymbol{a}_X$ over $\mathbb{X}$ representing the prior probability.
    \end{definition}
    In addition, a projection of a multinomial opinion into a classical probability distribution is possible through the projected probability distribution
    \begin{equation} \label{eqn:projected-probability}
        \boldsymbol{P}_X(x) = \boldsymbol{b}_X(x) + \boldsymbol{a}_X(x) u_X, \quad \forall x \in \mathbb{X} ,
    \end{equation}  
    representing the expected outcome of an opinion in probability space.

    In order to combine multiple opinions from various sources of the same domain of interest $\mathbb{X}$, several SL fusion operators exist.
    Various fusion operators for different types of problems can be found in~\cite{josang2016subjective}.
    One of them is the \textit{aleatory cumulative belief fusion} (A-CBF), which is appropriate for our needs.
    \begin{definition}[Aleatory Cumulative Belief Fusion]
    	Let $\omega_X^A$ and $\omega_X^B$ be multinomial opinions of source $A$ and $B$ of the same variable $X \in \mathbb{X}$. 
    	Then, the fused opinion $\omega_X^{A \diamond B}$ is defined as
    	\begin{equation} \label{eqn:a-cbf}
        	\omega_X^{A \diamond B} = \left \{  \begin{array}{ll}
        	\boldsymbol{b}_X^{A \diamond B}(x) &= \frac{ \boldsymbol{b}_X^{A}(x) u_X^{B} + \boldsymbol{b}_X^{B}(x) u_X^{A} }{ u_X^{A} + u_X^{B} - u_X^{A} u_X^{B} } \\
        	\\[-6pt]
        	u_X^{A \diamond B} &= \frac{ u_X^{A} u_X^{B} }{ u_X^{A} + u_X^{B} - u_X^{A} u_X^{B} }\\
        	\\[-6pt]
        	\boldsymbol{a}_X^{A \diamond B}(x) &= \frac{ \boldsymbol{a}_X^{A}(x) u_X^{B} + \boldsymbol{a}_X^{B}(x) u_X^{A} }{ u_X^{A} + u_X^{B} - 2 u_X^{A} u_X^{B} } 
        	\\[6pt]
        	&\quad - \frac{ (\boldsymbol{a}_X^{A} (x) + \boldsymbol{a}_X^{B}(x)) u_X^{A} u_X^{B} }{ u_X^{A} + u_X^{B} - 2 u_X^{A} u_X^{B} } 	 	
    	\end{array}\right .
    	\end{equation}
    	for $u_X^A \neq 0$ $\lor$ $u_X^B \neq 0$ and $u_X^A \neq 1$ $\lor$ $u_X^B \neq 1$. 
    	The operator $\oplus$ for $\omega_X^{A \diamond B} = \omega_X^A \oplus \omega_X^B$ is called aleatory cumulative belief fusion. 
    	For corner cases of the definition, we refer to~\cite{josang2016subjective}.
    \end{definition}
    The reverse operator of cumulative belief fusion is called cumulative unfusion~\cite{josang2016subjective}.
    With this, a specific opinion is removed from an already fused opinion.
    \begin{definition}[Cumulative Unfusion]
    	Consider a fused opinion $\omega_X^C = \omega_X^{A \diamond B}$ as in~\eqref{eqn:a-cbf} for a known opinion $\omega_X^B$ of $X \in \mathbb{X}$ with the same base rate as the otherwise unknown opinion $\omega_X^A$, in fact $\boldsymbol{a}_X$.
    	Then, the unfused opinion $\omega_X^A = \omega_X^{C \bar{\diamond} B}$ is obtained by
    	\begin{equation} \label{eqn:cumulative-unfusion}
        	\omega_X^{C \bar{\diamond} B} = \left \{  \begin{array}{ll}
        	\boldsymbol{b}_X^{C \bar{\diamond} B}(x) &= \frac{ \boldsymbol{b}_X^{C}(x) u_X^{B} - \boldsymbol{b}_X^{B}(x) u_X^{C} }{ u_X^{B} - u_X^{C} + u_X^{B} u_X^{C} } \\
        	\\[-6pt]
        	u_X^{C \bar{\diamond} B} &= \frac{ u_X^{B} u_X^{C} }{ u_X^{B} - u_X^{C} + u_X^{B} u_X^{C} }\\
        	\\[-6pt]
        	\boldsymbol{a}_X^{C \bar{\diamond} B}(x) &= \boldsymbol{a}_X(x)	 	
        	\end{array}\right .
    	\end{equation}
    	for $u_X^B \neq 0$ $\lor$ $u_X^C \neq 0$. 
    	The operator $\ominus$ in $\omega_X^{C \bar{\diamond} B} = \omega_X^C \ominus \omega_X^B$ is called cumulative belief unfusion. 
    	For the corner case $u_X^B = u_X^C= 0$, we refer to~\cite{josang2016subjective}.
    \end{definition}
    In general, for trust and belief modeling of transitive trust paths, the trust discounting operator is used~\cite{josang2016subjective}.
    We use trust discounting slightly differently regarding our purpose.
    In fact, we apply the operator in connection with the estimation of time-varying parameters to account for information degradation over time due to possible parameter changes.
    \begin{definition}[Trust Discounting] 
    	Consider opinion $\omega_X^A$ over $X \in \mathbb{X}$ and a discount probability $p_{td} \in [0, 1]$.
    	Then, the trust discounted opinion $\omega_X^{A_{p_{td}}} =\emph{TD}\left(\omega_X^{A}, p_{td}\right)$ is given by 
    	\begin{equation} \label{eqn:TrustDiscounting}
        	\omega_X^{A_{p_{td}}} = \left \{ \begin{array}{ll}
        	\boldsymbol{b}_X^{A_{p_{td}}}(x) &= p_{td} \; \boldsymbol{b}_X^{A}(x) \\
        	u_X^{A_{p_{td}}} &= 1 - p_{td} \sum \limits_{x \in \mathbb{X}} \!  \boldsymbol{b}_X^{A}(x) \\
        	\boldsymbol{a}_X^{A_{p_{td}}}(x) &= \boldsymbol{a}_X^{A}(x) 
        	\end{array}\right . 
    	\end{equation}
    	with $\emph{TD}\left(\omega_X^{A}, p_{td}\right)$ denoting trust discounting of opinion $\omega_X^{A}$ with respect to $p_{td}$.
    \end{definition}
    To compare two different opinions, the \textit{degree of conflict} (DC) is presented to measure the difference between opinions about the same variable $X \in \mathbb{X}$.
    \begin{definition}[Degree of Conflict]
    	Consider two multinomial opinions $\omega_X^A$ and $\omega_X^B$ of source $A$ and $B$ over $X \in \mathbb{X}$. 
    	Then, $\emph{DC} \left(\omega_X^A, \omega_X^B\right)$ denotes the degree of conflict between $\omega_X^A$ and $\omega_X^B$, which is defined as
    	\begin{equation} \label{eqn:DC} 
    		\emph{DC} \left(\omega_X^A, \omega_X^B\right) = \emph{PD} \left(\omega_X^A, \omega_X^B\right) \cdot \emph{CC} \left(\omega_X^A, \omega_X^B\right),
    	\end{equation} 
    	consisting of the projected distance
    	\begin{equation} \label{eqn:PD}
    	    \emph{PD} \left(\omega_X^A, \omega_X^B\right) = \frac{1}{2} \sum_{x \in \mathbb{X}} |\boldsymbol{P}_{X}^A (x) - \boldsymbol{P}_{X}^B (x)| 
    	\end{equation}   
        and the conjunctive certainty
        \begin{equation} \label{eqn:CC}
            \emph{CC} \left(\omega_X^A, \omega_X^B\right) = \left( 1 - u_X^A\right) \left( 1 - u_X^B\right) . 
        \end{equation} 
    \end{definition}
    Thus $\text{DC} \in \left[0, 1\right]$ holds, and we can say more precisely that the smaller the $\text{DC}$ is, the more similar the opinions are, and vice versa.

    \section{Self-Assessment for Kalman Filtering} \label{section:self_assessment_sot_kalman}
    
    This section briefly describes our basic algorithm for the self-assessing Kalman filter with SL introduced in~\cite{griebel2020kalman}.
    However, on top of that, we present a novel threshold for the comparison measure DC in the context of our SA module.

    \subsection{Basic Algorithm from~\cite{griebel2020kalman}}
    \label{section:self_assessment_sot_kalman_basic_algo}
    
    For every measurement, we generate an SL multinomial opinion $\omega_X^{z}$ 
    about the correctness of the Kalman filter’s assumptions based on information from the measurement prediction.
    In addition to the opinion generated in each step, there are two other opinions, $\omega_X^{st}$ and $\omega_X^{lt}$, respectively representing a short-term and long-term memory.
    Each newly generated opinion $\omega_X^{z}$ of incoming measurements is first fused to the short-term opinion $\omega_X^{st}$ using the A-CBF~\eqref{eqn:a-cbf}.
    If a given amount of evidence $n_{st} \in \mathbb{N}$ has been accumulated in $\omega_X^{st}$, we compare $\omega_X^{st}$ and the long-term opinion $\omega_X^{lt}$ using the DC~\eqref{eqn:DC}.
    When a conflict is identified, meaning the DC exceeds a threshold, then $\omega_X^{lt}$ is reset and overwritten with $\omega_X^{st}$.
    Otherwise, the two opinions are fused into $\omega_X^{lt}$ to accumulate evidence. 
    Besides, $\omega_X^{st}$ continuously represents the opinion over the last $n_{st} \in \mathbb{N}$ time steps.
    This means that at each time step, a new opinion of the incoming measurement is fused using the A-CBF, and the opinion of the oldest included measurement is unfused using cumulative unfusion~\eqref{eqn:cumulative-unfusion}.
    In addition, trust discounting~\eqref{eqn:TrustDiscounting} is used to model the degradation of information about the measurements over time.
    Lastly, to obtain an SA score, we compare $\omega_X^{lt}$ with a reference opinion $\omega_X^{ref}$, which models the filter assumptions.
    For this, the DC is used.
    This gives us an SA online measure with an associated uncertainty in each time step. 
    Comparisons based on the DC require a threshold that identifies a significant violation of the assumptions and a resulting conflict.
    The threshold was an adjustable parameter so far that had to be selected manually.
    For detailed information on the algorithm, we refer the reader to our work in~\cite{griebel2020kalman}.
    
    \subsection{Degree of Conflict Threshold Derivation} \label{section:dc-threshold-derivation}
    
    In this work, we extend the SA module of~\cite{griebel2020kalman}, which was summarized in the previous subsection, with a threshold derivation for the SL comparison measure DC.
    This threshold denotes a limit where the difference between two opinions exceeds a certain confidence level.
    In our application, this means that, by exceeding the threshold, we expect that assumptions made in the tracking algorithm are violated.
    
    We want to detect whether an SL opinion differs from a reference opinion with a certain confidence level.
    To this end, we consider multinomial distributions because the domain of opinions with exclusive events can be represented as a multinomial distribution.
    For the event probabilities of a multinomial distribution $p_1, \ldots, p_W \in [0, 1]$ with $W \in \mathbb{N}$, it holds that $\sum_{i=1}^W p_i = 1$.
    Given samples and a requested confidence, we calculate the limits of a corresponding confidence interval and, thus, a threshold for the DC.
    
    We base on the approach from~\cite{thompson1987sample}; however, we make some modifications: 
    We calculate the confidence interval instead of the sample size and consider the counter event.
    The latter is given by
    \begin{align}
        \forall p_i, \pi_i \in [0, 1], i = 1, \ldots W, \alpha \in [0, 1]: \nonumber\\
        \exists d \in \mathbb{R}, p_i \not\in [\pi_i - d, \pi_i + d ] ,
    \end{align}
    which means that with a confidence $\alpha \in [0, 1]$ the probability $p_i \in [0, 1]$ is outside the confidence interval around the reference probability $\pi_i \in [0, 1]$ with calculated limit $d \in \mathbb{R}$.
    Due to the counter event, the consideration of the worst-case scenario like in~\cite{thompson1987sample} is not necessary anymore.
    And the confidence $\alpha \in [0, 1]$ corresponds to an error probability in this case.
    Applying the formulas from~\cite{thompson1987sample}, we consequently obtain the limit $d \in \mathbb{R}$ of the confidence interval $[\pi_i - d, \pi_i + d ]$ depending on $\alpha$ as follows:
    \begin{align}
        d \left( \alpha \right) &= \sqrt{\frac{d^2 n_s}{n_s}}
    \end{align}
    with 
    \begin{align}
        d^2 n_s &= \left[ \Phi^{-1} \left(1 - \frac{1 - \alpha}{2 W} \right) \right]^2 \frac{W-1}{W^2} \, ,
    \end{align} 
    where $W \in \mathbb{N}$ is the number of events in the domain $\mathbb{X}$, 
    ${n_s \in \mathbb{N}}$ the sample size, 
    and $\Phi^{-1}$ the inverse Gaussian distribution.
    Here, it is assumed that the assumptions of~\cite{thompson1987sample} are satisfied, especially for the normal approximation where $n_s$ should be large enough. 
    However, this is reasonable due to the parameter choice of $n_{st}$ in~\cite{griebel2020kalman}.
    
    Let $\omega_X^A$ with $u_X^A=0$ be the reference opinion and $\omega_X^B$ with $0 < u_X^B < 1$ be an evidence-based opinion to verify the assumed reference, with corresponding projected probabilities~\eqref{eqn:projected-probability} $\boldsymbol{P}_{X}^A$ and $\boldsymbol{P}_{X}^B$, respectively.
    To compute the threshold for the DC, opinion $\omega_X^B$ is defined such that each component of the projected probabilities differs from the corresponding reference by at most the limit $d$, namely $\boldsymbol{P}_{X}^B (x) = \boldsymbol{P}_{X}^A (x) \pm d , \; \forall x \in \mathbb{X}$.
    Using this relation, the bijective mapping theorem from~\cite{josang2016subjective} between multinomial opinions and Dirichlet probability density functions, and the definition of the DC in~\eqref{eqn:DC}, we obtain the threshold
    \begin{align}
        \eta \left( \alpha \right) &= \text{DC} \left(\omega_X^A, \omega_X^B \right) = 
        \frac{1}{2} \left( 1 - u_X^B\right) \sum_{x \in \mathbb{X}} d \left( \alpha \right)  \nonumber\\
        &= \frac{1}{2} W d \left( \alpha \right) \left( 1 - \frac{W}{W + n_s^B} \right) ,
    \end{align} 
    where $n_s^B$ the amount of evidence of $\omega_X^B$.
    All in all, the requested error probability $\alpha \in [0, 1]$ is the only design parameter here.
    All other parameters are given by the opinion setup and the received data.
    
    A trade-off has to be made between false positives and false negatives to find a suitable value for the error probability $\alpha$.
    Choosing a high value for $\alpha$ increases the number of false positives and, thus, triggers a reaction to situations where no reaction would be required.
    On the other side, choosing a low value for $\alpha$ will result in a higher false-negative rate and missing reactions to errors that actually occurred.
    Thus, choosing a higher error probability $\alpha$ is usually better.
    As a result, fewer errors are overlooked, and a more thorough review of the results takes place more often, even when it would not be necessary.

    \section{Self-Assessment for Single-Object Tracking in Clutter} 
    \label{section:self_assessment_sot_clutter}

    This section extends SOT by the clutter aspect.
    Starting with the problem formulation, we mathematically derive the reference distribution for our SA regarding our studied problem.
    Accordingly, we present the resulting SA module for SOT in clutter. 

    \subsection{Problem Formulation}
    
    Our proposed method aims to perform an online SA for SOT in clutter using the Kalman filter framework and the nearest neighbor data association algorithm. 
    To this end, we want to monitor the overall performance of the filter algorithm and all statistical assumptions online. 
    These are the detection probability $p_D(x) \in [0,1]$, the clutter rate $m^c \in \mathbb{N}_0$, and the spatial uncertainty of the measurements $z \in \mathbb{R}^{m_z}$.
    Additionally, we aim to obtain explicit uncertainty information that expresses the reliability of these measures in terms of statistical evidence.

    \subsection{Theoretical Derivation}
    
    Starting from the measurement likelihood in~\eqref{eqn:measurment-likelihood}, 
    we first apply the following tracking assumptions to end up with a closed-form expression, namely, that the detection probability $p_D(x)$ is constant, i.e., $p_D(x) = p_D$ is independent of state $x$.
    In addition, we make the classical Kalman filter assumptions that all object signals and probability densities are Gaussian distributed and the models are linear.
    This also means explicitly that the measurement noise is Gaussian distributed.
    Moreover, clutter measurements are assumed to be uniformly distributed over the sensor FOV, i.e., $\lambda_c (z) = \lambda_c = \bar{\lambda}_c / \text{Vol}(\mathcal{R})$ for $z \in \mathcal{R}$, where $\mathcal{R}$ is the sensor FOV with the volume $\text{Vol}(\mathcal{R})$.
    Using these assumptions and considering the measurement likelihood on the sensor FOV $\mathcal{R}$ at time step $k \in \mathbb{N}$, we obtain
    \begin{align}
        p(Z_k | x_k) =& \left[ (1-p_D) +  \frac{p_D}{\lambda_c} \sum_{\theta_k=1}^{m_k} g_k (z^{\theta_k} | x_k) \right] \nonumber\\
        & \times \frac{\exp{(-\bar{\lambda}_c}) \lambda_c^{m_k}}{m_k!} \\
        =& \; c_0 + c_1 \sum_{\theta_k=1}^{m_k} g_k \left( z^{\theta_k} | x \right) \label{eqn:measurement-likelihood-simple1}
    \end{align} 
    with the two parameters 
    \begin{align}
        c_0 \left( p_D, \lambda_c, m_k \right) &= \left( 1-p_D \right) \frac{\exp{(-\bar{\lambda}_c})}{m_k!} \lambda_c^{m_k} , \\
        c_1 \left( p_D, \lambda_c, m_k \right) &= p_D \frac{\exp{(-\bar{\lambda}_c})}{m_k!} \lambda_c^{m_k-1} .
    \end{align}
    Considering the likelihood function $g_k \left( z^{\theta_k} | x_k \right)$ in~\eqref{eqn:measurement-likelihood-simple1}, we sum over all data association hypotheses of the measurements.
    To make all data association hypotheses in~\eqref{eqn:measurement-likelihood-simple1} comparable, we use the Gaussian distributed measurement prediction $z_{k+1|k} \sim \mathcal{N} \left(h (\hat{x}_{k+1|k}), S_k \right)$ from Kalman filtering, see, e.g.,~\cite{bar1988tracking,bar2001estimation}.
    Then, we transform the input $z^{\theta_k}$ of the likelihood $g_k \left( z^{\theta_k} | x_k \right)$, namely the incoming measurements, using the residual
    \begin{align}
        \gamma^{\theta_k} = z^{\theta_k} - h (\hat{x}_{k+1|k}), \quad \theta_k = 1, \ldots, m_k 
    \end{align}
    with the corresponding squared Mahalanobis distance for normalization purposes to obtain the transformed measurement
    \begin{align}
        \tilde{z}^{\theta_k} = \left( \gamma^{\theta_k} \right)^T S_k^{-1} \gamma^{\theta_k}, \quad \theta_k = 1, \ldots, m_k .
    \end{align}
    Using the transformed measurement as input, we obtain the transformed measurement likelihood 
    \begin{align}
        g_k \left( \tilde{z}^{\theta_k} | x_k \right) \sim \chi^2_{m_z} , \quad \theta_k = 1, \ldots, m_k \label{eqn:measurement-likelihood-transformed}
    \end{align}
    that follows a $\chi^2_{m_z}$ distribution with $m_z$ degrees of freedom, corresponding to the dimension of each measurement.
    This means that the distribution of the transformed measurement likelihoods of all data association hypotheses only depends on the dimension of the measurements $m_z$, which is the same for all measurements in all time steps.
    Thus, the transformed measurement likelihoods of all data association hypotheses can be gathered in one distribution for all measurements of a single time step $k \in \mathbb{N}$.
    To this end, we obtain
    \begin{align}
        p(\tilde{Z}_k | x_k) &= c_0 + c_1 \sum_{\theta_k=1}^{m_k} g_k \left( \tilde{z}^{\theta_k} | x \right) \\
        &\sim c_0 + c_1 m_k \, \chi^2_{m_z} . \label{eqn:measurement-likelihood-simple2}
    \end{align}
    Because the distribution of the transformed measurement likelihood in~\eqref{eqn:measurement-likelihood-transformed} is additionally independent of time, we can compare all measurements of all time steps with one theoretical reference distribution.
    We, therefore, consider the expected value of the number of measurement $m_k$ at any time step for a long-term average to obtain a closed-form expression of the reference distribution.
    Consequently, this results in the expected value $\mathbb{E}[ m_k ] = \bar{\lambda}_c + p_D$.
    All in all, we end up with the final representation for the reference distribution of the transformed measurement likelihood: 
    \begin{align}
        p(\tilde{Z} | x) &\sim
        \tilde{c}_0 + c_1 (\bar{\lambda}_c + p_D) \, \chi^2_{m_z} \\
        &\sim \tilde{c}_0 + \tilde{c}_1 \, \chi^2_{m_z} \label{eqn:measurement-likelihood-transformed-final}
    \end{align}
    with
    \begin{align}
        \tilde{c}_0 \left( p_D, \lambda_c \right) &= \left( 1-p_D \right) \frac{\exp{(-\bar{\lambda}_c})}{ \left( \bar{\lambda}_c + p_D \right) !} \lambda_c^{ \left( \bar{\lambda}_c + p_D \right) } , \\
        \tilde{c}_1 \left( p_D, \lambda_c \right) &= p_D \frac{\exp{(-\bar{\lambda}_c})}{\left( \bar{\lambda}_c + p_D \right)!} \lambda_c^{\left(\bar{\lambda}_c + p_D-1\right)} \left( \bar{\lambda}_c + p_D \right) ,
    \end{align}
    where we define and extend the factorial function to non-integer arguments $x \in \mathbb{R}$ by using the interpolation relation with the Gamma function $(x)! = \Gamma (x+1)$.
    
    When considering the corner case $p_D = 1$ and $\bar{\lambda}_c = 0$, it follows that $\lambda_c = 0$, and we get $\tilde{c}_0 (1, 0) = 0$ and $\tilde{c}_1 (1, 0) = 1$.
    Thus, we end up with the transformed measurement likelihood for classical Kalman filtering $p(\tilde{Z} | x) = g \left( \tilde{z} | x \right) \sim \chi^2_{m_z}$, which, in fact, corresponds to the classical consistency measure in Kalman filtering, the NIS.

    \subsection{Self-Assessment Module}

    Our proposed SA module for SOT in clutter using nearest neighbor association consists of four different components.
    First, the overall opinion is described, where all assumptions are joined and checked as a whole. 
    Afterward, the three different components, i.e., association, measurement, and clutter, are checked individually.
    For all checks in the SA module, we use the general approach of our SL-based distribution test from~\cite{griebel2020kalman}, which is briefly described in Section~\ref{section:self_assessment_sot_kalman_basic_algo}.
    
    \subsubsection{Overall Opinion}
    
    With the overall opinion, the SA can make a general statement of the tracking quality.
    On the one hand, a simply tangible measure for tracking assessment is obtained.
    On the other hand, when tracking performs poorly, it is unclear why and especially which assumption is violated. 
    For calculation of the overall opinion, the transformed measurement likelihood~\eqref{eqn:measurement-likelihood-transformed-final} is used as reference.
    This reference distribution is modeled and implemented as the SL opinion $\omega_{X, overall}^{ref}$, designed as a dogmatic opinion with $u_X = 0$.
    To check $\omega_{X, overall}^{ref}$ and obtain the overall SA score, we apply our SL-based distribution test using the general approach in Section~\ref{section:self_assessment_sot_kalman_basic_algo} and accumulate and represent the collected data in a long-term SL opinion $\omega_{X, overall}$.
    
    \subsubsection{Association Opinion}
    
    The second opinion focuses on the association component mainly influenced by the detection probability $p_D \in [0,1]$.
    The underlying opinion is constructed as binomial opinion with the event that an association has or has not taken place.
    Starting from~\eqref{eqn:measurement-likelihood-transformed-final}, we split the likelihood into two parts, where
    $\tilde{c}_0 \left( p_D, \lambda_c \right)$ displays the reference probability that no association took place 
    and the probability resulting by integrating $\tilde{c}_1 \left( p_D, \lambda_c \right) \, \chi^2_{m_z}$ over the corresponding limits represents the reference probability that an association took place.
    This results in our reference opinion $\omega_{X, assoc}^{ref}$.
    By using the nearest neighbor association, we can accurately determine in each time step whether or not an association has taken place and generate a corresponding opinion with evidence of one.
    This opinion is fused with our evidence-based long-term opinion $\omega_{X, assoc}$ to check the reference opinion $\omega_{X, assoc}^{ref}$ using the general approach of our SL-based distribution test.

    \subsubsection{Measurement Opinion}

    The following opinion represents the spatial uncertainty of the associated measurement.
    Starting from the overall opinion and the transformed measurement likelihood in~\eqref{eqn:measurement-likelihood-transformed-final}, we focus on the second part $\tilde{c}_1 \left( p_D, \lambda_c \right) \, \chi^2_{m_z}$ for the quality measure, which contains the event that the object is detected.
    The opinion generation of $\omega_{X, meas}^{ref}$ and $\omega_{X, meas}$ works similarly to the overall opinion, except that for a missed detection, i.e., no measurement is associated, we generate a vacuous opinion. 
    This means that, in this case, we have the information that no statement can be made, which results in no belief and total uncertainty.
    
    \subsubsection{Clutter Rate Opinion}
    
    The opinion $\omega_{X, clutter}$ checks the assumed Poisson distributed clutter rate $m_k^c \sim \text{Poi} \left( \bar{\lambda}_c \right)$, modeled by the reference $\omega_{X, clutter}^{ref}$.
    This opinion is entirely independent of the other opinions, especially of the overall opinion.
    Moreover, the spatial distribution of clutter is not considered in this opinion.
    We model this opinion based on confidence intervals of the Poisson distribution, namely, $50 \%$ confidence around the expected value and $25 \%$ below and above this middle confidence interval.

    \section{Simulations}\label{section:simulations}

    \begin{figure*}[!t]
	    \center
	    \input{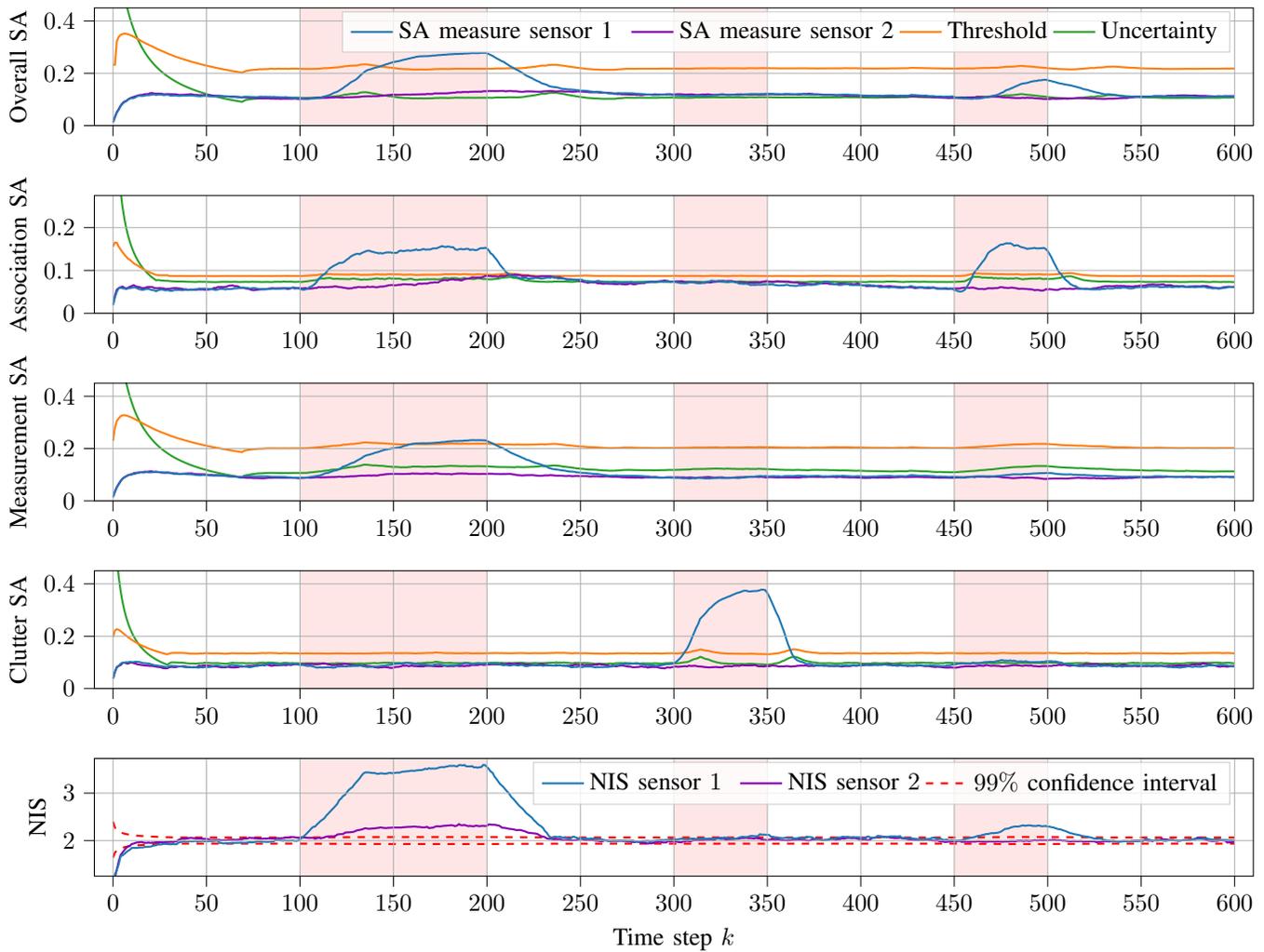}
	    \caption{Results of our SA module in a challenging simulation scenario where the measurement noise, clutter rate, and detection probability of sensor $1$ are consecutively disturbed, which is indicated with red shadow areas. 
	    Our subjective logic-based SA module, consisting of an overall, association, measurement, and clutter SA, is compared to the time-average NIS. The results are averaged values from $200$ Monte Carlo runs.
	    \label{fig:simulation_sceanrio}}
    \end{figure*}
    We construct a challenging simulation scenario to evaluate our proposed approach, which can be motivated by, e.g., adverse weather conditions in an automated driving scenario.
    In the scenario, we consider a single-object multi-sensor simulation setup where three sensors, measuring the position in two dimensions $(x,y) \in \mathbb{R}^2$, are used to track a single object in clutter.
    In the beginning, all sensors are modeled the same for the tracking assumptions, i.e., the detection probability, the clutter process, and the spatial uncertainty.
    The single object being tracked may be, for example, a vehicle traveling in front of an automated vehicle that is to be followed using adaptive cruise control.
    
    In the simulation scenario, sensor $1$ is our disturbed sensor while the others are still functioning normally.
    This can be motivated by the fact that sensor $1$ is a particular type of sensor that is more susceptible to adverse weather conditions, such as a camera.
    In the time steps $100$ to $200$, the ground truth spatial measurement noise of sensor $1$ is increased from $\sigma_w = \SI{0.75}{m}$ by a factor of $4$, while the assumed noise in our tracking algorithm is kept constant.
    This corresponds to a degradation of the sensor accuracy, e.g., due to fog, affecting the view of the camera sensor.
    Then, from time step $300$ to $350$, the expected ground truth clutter rate of sensor $1$ is decreased from $\bar{\lambda}_c = 4$ by half, while the assumed rate in our tracking algorithm remains constant.
    At first glance, this sounds counterintuitive because tracking usually gets better when there is less clutter.
    However, this is the more challenging case for the SA module since a higher clutter rate possibly leads to false clutter associations and, thus, to track divergences.
    These effects can be detected by the SA and lead to a significant increase in SA scores for all opinions.
    A decreasing clutter rate is much harder to detect and can be caused, for example,
    by an improvement of the conditions in changeable weather, which can result in fewer clutter detections in the camera sensor.
    Finally, from time step $450$ to $500$, the ground truth object detection probability of sensor $1$ is reduced from $p_D = 0.9$ to $\tilde{p}_D = 0.6$, while the assumed probability in our tracking algorithm remains constant.
    This means that the sensor has more difficulty detecting the object, which can be caused, for example, by the camera being blinded at sunset.
    
    The simulation results, which are averaged values from $200$ Monte Carlo runs, are visualized in Fig.~\ref{fig:simulation_sceanrio}.
    Our SL-based SA module is compared to the time-average NIS.
    Moreover, we plot the scores for the disturbed sensor $1$ and, as an exemplary comparison for a normally working sensor, for sensor $2$.
    The NIS is visualized along with the $99 \%$ confidence interval calculated under the influence of the time-average version of the NIS and the Monte Carlo runs.
    In comparison, our SL-based SA scores, including the sensor measures and the corresponding explicit uncertainty, are presented along with the dynamically calculated threshold for the DC, which we derived in Section~\ref{section:dc-threshold-derivation}.
    
    It can be seen that after time step $100$, the sensor $1$ SA scores of the overall, association, and measurement SA rise above the threshold, indicating that the reliability of the tracking has deteriorated.
    The association SA reacts the fastest, followed by the overall and the measurement SA.
    This is because it is pretty challenging to identify the spatial measurement uncertainty of a sensor in the presence of clutter.
    In contrast, the time-average NIS reacts even faster and quickly exceeds the confidence interval.
    However, the NIS score of sensor $2$, which operates normally, also exceeds the confidence interval.
    Here, the SL-based SA scores correctly reflect the behavior of sensor $2$ and are below the thresholds.
    At time step $300$, the clutter SA score of sensor $1$ rises rapidly above the threshold.
    In fact, this is the only SA score that clearly detects the decreased clutter rate.
    The time-average NIS does not detect this aspect.
    As mentioned earlier, while a lower clutter rate generally improves the tracking, knowledge of this rate and further adjustments to the parameter can improve the tracking results. 
    Then, at times step $450$, the detection probability of sensor $1$ is reduced, which is
    slightly noticeable in the overall SA score, without exceeding the threshold.
    However, the association SA quickly and clearly detects the assumption violation.
    The time-average NIS, on the other hand, increases only gradually.
    
    All in all, this shows that our proposed SA is capable of detecting the respective malfunctions and indicating them accordingly.
    The SL-based SA exhibits a more detailed and focused analysis than the time-average NIS.
    Moreover, it is shown that the NIS-based SA fails to detect a decreased clutter rate and wrongly detects failures also in other sensors, while the proposed approach works as expected.

    \section{Conclusion} \label{section:conclusion}
    
    In this work, we proposed a novel SL-based SA scheme for SOT in clutter with an overall and component analysis of certain tracking assumptions.
    The online reliability scores of our method include a measure for the statistical uncertainty to gain a comprehensive overall assessment.
    We mathematically derived a formula to adequately choose the sensitivity threshold for the comparison measure DC.
    Furthermore, the reference distribution for our SA module was theoretically derived.
    A challenging simulation scenario showed that the proposed approach reliably detects sensor accuracy degeneration.
    Furthermore, our SA can also clearly identify wrong assumptions on the clutter rate or object detection probability.
    At the same time, the NIS, as the state-of-the-art baseline method, has its difficulties.
    
    In future work, we want to extend our SA to further association algorithms for SOT in clutter, e.g., the probabilistic data association.
    In this context, we want to build up a general concept and structure of SA modules with interchangeable association algorithms for SOT in clutter.
    Furthermore, we want to extend our approach to multi-object tracking.

    \bibliographystyle{IEEEtran}
    \bibliography{references}

\end{document}